\documentclass[twocolumn]{aastex631}
\usepackage{amsmath}

\begin{document}

\title{Constraining the equation of state of hybrid stars\\ using recent information from multidisciplinary physics}

\correspondingauthor{Debarati Chatterjee}
\email{debarati@iucaa.in}

\author[0000-0001-8604-5362]{Swarnim Shirke}
\affiliation{Inter-University Centre for Astronomy and Astrophysics, \\Post Bag 4, Ganeshkhind, Pune 411 007, India}

\author[0000-0002-1656-9870]{Suprovo Ghosh}
\affiliation{Inter-University Centre for Astronomy and Astrophysics, \\Post Bag 4, Ganeshkhind, Pune 411 007, India}

\author[0000-0002-0995-2329]{Debarati Chatterjee}
\affiliation{Inter-University Centre for Astronomy and Astrophysics, \\Post Bag 4, Ganeshkhind, Pune 411 007, India}

\begin{abstract}

At the ultra-high densities existing in the core of neutron stars, it is expected that a phase transition from baryonic to deconfined quark matter may occur. Such a phase transition would affect the underlying equation of state (EoS) as well as the observable astrophysical properties of neutron stars. Comparison of EoS model predictions with astronomical data from multi-messenger signals then provides us an opportunity to probe the behaviour of dense matter. In this work, we restrict the allowed parameter space of EoS models in neutron stars for both nucleonic (relativistic mean field model) and quark matter (bag model) sectors by imposing state-of-the-art constraints from nuclear calculations, multi-messenger astrophysical data and perturbative QCD (pQCD). We systematically investigate the effect of each constraint on the parameter space of uncertainties using a cut-off filter scheme, as well as the correlations among the parameters and with neutron star astrophysical observables. Using the constraints, we obtain limits for maximum NS mass, maximum central density, as well as for NS radii and tidal deformability. Although pQCD constraints are only effective at very high densities, they significantly reduce the parameter space of the quark model. We also conclude that astrophysical data supports high values of the bag parameter $B$ and disfavors the existence of a pure quark matter core in hybrid stars.

\end{abstract}


\section{Introduction} \label{sec:intro}

One of the most intriguing questions in physics is the fundamental constitution of matter. High-energy experiments are performed in nuclear laboratories, as well as heavy-ion collision experiments in accelerators to probe the nature of matter under extreme conditions of temperature and density, which explore different regions of the phase diagram of Quantum Chromodynamics (QCD), the theory of strong interactions \citep{Baym2018quarkreviewads, blaschke2020book}. Despite recent advances\citep{fodor2004latticeqcdads, Aoki2006latticeqcdads, Bazavov2017lattiqcdads, Bazavov2019latticeqcdads}, lattice QCD calculations are not applicable at finite chemical potentials due to the sign problem \citep{forcrand2010qcdsignproblem}. Theoretical calculations from Chiral Effective Field Theory (CEFT) \citep{drischler2016ceftads, drischler2019} for low density nuclear matter and perturbative QCD (pQCD) \citep{kurkela2010pqcd, gorda2018pqcdads} for high density quark matter provide some reliable limits.

An alternative means to approach this problem is through astrophysical systems such as compact stars. Neutron Stars (NSs) span a wide range of densities, from nuclei at the surface to several times normal nuclear matter density in the core. NSs provide a natural environment for the appearance of strangeness-containing matter, such as hyperons or deconfined quark matter in its interior, due to the ultrahigh densities prevailing there \citep{lattimerprakash2004ads, Glendenning1997book}. Therefore modeling NSs provides an efficient way of probing hadron-quark phase transition in an astrophysical environment. 

It is conjectured that if a phase transition from hadronic to quark matter (crossover or first order) occurs in the NS interior, it could significantly affect several observable NS properties, such as its mass, radius, moment of inertia, or even gravitational wave (GW) emission \citep{alford2007, Bauswein2019PostMergerHybrid, alford2019signaturesads}. NSs have been observed since many decades at multiple frequencies across the electromagnetic spectrum, from which one can deduce important global properties such as its mass or radius, which can be related to its EoS. Precise measurements of the NS maximum mass \citep{demorest2010maxmassads, antoniadis2013maxmassads, linares2018maxmassads, linares2020maxmassads, cromartie2020maxmassads, fonseca2021ApJmaxmassads} impose stringent constraints on the NS core composition. Although radius measurements from quiescent low mass X-ray binaries and thermonuclear bursts of accreting NSs suffer from larger uncertainties \citep{guillot2013radiusads, ozel2016radiusmeasurementNICERads, ozel2016massradiusads}, the NICER mission recently launched by NASA is providing radius estimates with much higher accuracy \citep{miller2019nicermrj0030ads, riley2019nicermrj0030ads, miller2021nicermrj0740ads, riley2021nicermrj0740ads}. Further, a breakthrough has emerged with the recent direct detection of gravitational waves by the LIGO-Virgo Collaboration \citep{Abbott2017AGW170817, abbott2017BGW170817multi}. In particular, the detection of GWs from the binary NS merger GW170817  and the inferred tidal deformability of the components NSs in the inspiral phase has led to important implications for the equation of state (EoS) of dense matter \citep{Abbott2017AGW170817, abbott2017BGW170817multi, abbottetal2019properties}. NS multi-messenger (MM) observations can therefore provide a wealth of information about NS internal composition \citep{bauswein2017radiusfromgw, margalit2017, annala2018gwconstraints, Most2018radiuslamfromgw, fattoyev2018neutronskinfromgw, Paschalidis2018ILoveQHybrid, radice2018, rezzolla2018ads}.

It was shown that imposing complementary limiting constraints at high densities from pQCD in addition to low-density nuclear matter from CEFT reduces the uncertainty of NS matter significantly \citep{kurkela2014constraining, Komoltsev2022}. The highest observed NS mass $\sim$ 2 $M_{\odot}$ \citep{demorest2010maxmassads, antoniadis2013maxmassads, cromartie2020maxmassads, fonseca2021ApJmaxmassads} further imposes important consequences on the speed of sound for a given EoS. Combining additional constraints from tidal deformability dramatically reduces the allowed parameter space of EoSs \citep{annala2018gwconstraints}. In such studies, EoSs employed to interpolate between the low and high-density limits were piecewise polytropes or Chebyshev polynomials. Combining astrophysical observations with theoretical ab-initio calculations in a model-independent way, the behavior of the speed of sound was used to establish evidence of the presence of quark matter cores in NSs \citep{annala2020evidence}. Other multi-messenger NS data, e.g., measurements of pulsar radii from X-ray \citep{bogdanov2016, nattila2017} or NICER observations \citep{ riley2019nicermrj0030ads, miller2019nicermrj0030ads, miller2021nicermrj0740ads, riley2021nicermrj0740ads} has also been employed to impose strong constraints on the EoS and to test the presence of quark matter in NSs \citep{annala2022}.

The effect of phase transition in supranuclear matter on the GW signal is being studied with great interest \citep{pang2020parameterestimationPTads, chatziioannouhan2020}. Over the past couple of years, several works have used a Bayesian statistical approach to incorporate prior knowledge of multi-messenger data to perform a joint analysis of GW signal and its electromagnetic counterparts, combined with X-ray and radio pulsar observations and nuclear-theory calculations \citep{dietrich2020, coughlin2019, radice2019, raaijmakers2020}. Such studies have also been extended to investigate quark stars \citep{zhou2018, Li2020} or hybrid stars \citep{pang2021, Xie2021}. This is usually done by employing piecewise-polytropes, or constant speed-of-sound parametrization \citep{Alford2013CSS}. The drawback of such a scheme is that the constrained parameters (e.g., the speed of sound or polytropic index) cannot be directly linked to the properties of the microscopic quark matter EoS or the correlations of the underlying hadronic or quark parameters with global NS observables cannot be investigated.

A few recent studies considered hybrid stars with a phase transition from hadronic models based on nuclear interactions (such as RMF or CEFT) to quark matter (MIT Bag model) via a mixed phase \citep{parisi2021hybrid, nandi2018, nandi2021}. These works imposed constraints from NS maximum mass and tidal deformability in the light of GW170817 to constrain the model parameters. However, such works only considered selected EoS models (several of which are now incompatible with recent astrophysical data), which do not necessarily span the allowed parameter space of the hadronic/quark EoSs.

In two recent publications~\citep{ghosh2022multi,ghosh2022multihyperon}, the authors of this paper (S. G. and D. C.) explored the parameter space of hadronic matter within the framework of the Relativistic Mean Field model allowed by present uncertainties compatible with state-of-the-art nuclear/hypernuclear experimental data. They applied a Bayesian-like scheme with a hard cut-off to constrain the parameter space using multi-physics constraints at different density regimes: chiral effective field theory (CEFT), nuclear and heavy-ion collision data, as well as multi-messenger astrophysical observations of neutron stars. Using the posterior distributions, they investigated possible correlations between nuclear, hypernuclear, and astrophysical observables. This work extends this scheme to the quark degrees of freedom. By varying the parameters of the hadronic (RMF) and quark (MIT Bag) models within their present uncertainties, constraints are imposed using CEFT at low densities, multi-messenger astrophysical data at high densities, and pQCD at very high densities. We then investigate possible correlations among the model parameters as well as with NS global observables.

The paper is structured as follows: in Sec.~\ref{sec:formalism}, we outline the formalism used to carry out this study. In particular, we explain the microscopics of the generated EoS in Sec~\ref{sec:micro}, the global properties of NSs in Sec.~\ref{sec:macro}, and the details of the Bayesian scheme followed and constraints applied in Sec.~\ref{sec:bayesian}. We present the results of the works in Sec.~\ref{sec:results}. The final section, Sec.~\ref{sec:discussions}, discusses the implications of these results and important findings.

\section{Formalism} \label{sec:formalism}

\subsection{Microscopic description} \label{sec:micro}
There are different approaches to describing dense nuclear matter in NSs \citep{oertel2017}. While the ab-initio techniques employ calculations of nucleon-nucleon two-body interactions, the density functional models apply phenomenological techniques by adjusting model parameters to reproduce experimental nuclear observables. Similarly, there exist different descriptions for the pure quark matter phase, effective theories such as the Nambu-Jona-Lasinio (NJL) model \citep{NJL1961a, NJL1961b} or phenomenological models such as the MIT Bag model \citep{chodosjaffe1974newbagmodel, farhijaffe1984strange}. In this work, a phenomenological Relativistic Mean Field (RMF) model is used to describe the pure nucleonic phase, while the pure quark phase is described using the phenomenological MIT Bag model \citep{chodosjaffe1974newbagmodel, chodos1974baryonsinbag, farhijaffe1984strange}. The mixed phase is constructed by applying Gibbs phase rules \citep{Glendenning1997book}. NS matter in all phases is subject to chemical beta equilibrium and charge neutrality constraints. These phases are elaborated on in the sections below.

\subsubsection{Pure nucleonic phase} \label{sec:nucleonic}
To solve for the EoS of finite nucleonic matter, we adopt the RMF model with nucleons ($N$), i.e., protons ($p$) and neutrons ($n$), their interactions given by the exchange of three meson fields:  scalar sigma ($\sigma$), vector omega ($\omega$) and iso-vector rho ($\rho$) \citep{hornick2018, Chen2014, Fattoyev2010, GlenMos1991}. Leptons ($l$) present in the matter (here electrons and muons) are free and non-interacting \citep{weber1999pulsars}. The interaction Lagrangian is

\begin{align}\label{interaction_lagrangian}
\mathcal{L}_{int} &= \sum_{N} \bar\psi_{N}\left[g_{\sigma}\sigma-g_{\omega}\gamma^{\mu}\omega_{\mu}-\frac{g_{\rho}}{2}\gamma^{\mu}\boldsymbol{\tau\cdot\rho}_{\mu}\right]\psi_{N} \nonumber \\
&-\frac{1}{3}bm_{N}(g_{\sigma}\sigma)^{3}-\frac{1}{4}c(g_{\sigma}\sigma)^{4} \nonumber \\ 
&+ \Lambda_{\omega}(g^{2}_{\rho}\boldsymbol{\rho^{\mu}\cdot\rho_{\mu}})(g^{2}_{\omega}\omega^{\nu}\omega_{\nu}) + \frac{\zeta}{4!}(g^{2}_{\omega}\omega^{\mu}\omega_{\mu})^{2}~,
\end{align}
where $\psi$ is the Dirac spinor for fermions, $m_{N}$ is the vacuum nucleon mass, $\{\gamma^{i}\}$ are the gamma matrices, $\boldsymbol{\tau}$ are Pauli matrices. $g_{\sigma}, g_{\omega}, g_{\rho}$ are meson-nucleon coupling constants. The scalar and vector self-interactions couplings are $b$, $c$, and $\zeta$, respectively, while $\Lambda_{\omega}$ is the vector-isovector interaction. In the mean-field approximation, the mesonic fields are replaced by their expectation values $\bar\sigma=\langle\sigma\rangle$, $\bar\omega=\langle\omega^{0}\rangle$, $\bar\rho=\langle\rho^{0}_{3}\rangle$ in the ground state. The field equations are solved to obtain the energy density ($\epsilon)$ \citep{hornick2018}:
{\small
\begin{align}
\epsilon&=\sum_{N}\frac{1}{8\pi^{2}}\left[k_{F_{N}}E^{3}_{F_{N}}+k^{3}_{F_{N}}E_{F_{N}}
-m^{*4}\ln\left(\frac{k_{F_{N}}+E_{F_{N}}}{m^{*}}\right)\right]
\nonumber \\ &+\frac{1}{2}m^{2}_{\sigma}\bar\sigma^{2}+ \frac{1}{2}m^{2}_{\omega}\bar\omega^{2}+\frac{1}{2}m^{2}_{\rho}\bar\rho^{2} + \frac{1}{3}bm_{N}(g_{\sigma}\bar\sigma)^{3} + \frac{1}{4}c(g_{\sigma}\bar\sigma)^{4} \nonumber \\ &+ 3\Lambda_{\omega}(g_{\rho}g_{\omega}\bar\rho\bar\omega)^{2} + \frac{\zeta}{8}(g_{\omega}\bar{\omega})^{4}~,  \label{energydensity}\\
\end{align}
}
where,
\begin{align}
E_{F_{N}} &= \sqrt{k_{F_{N}}^{2} + m^{*2}}~, \nonumber \\ 
m^{*}&=m_{N}-g_{\sigma}\sigma~, \nonumber \\
\mu_{N} &= E_{F_{N}} + g_{\omega}\bar{\omega} + \frac{g_{\rho}}{2}\tau_{3N}\bar{\rho}~. \nonumber 
\end{align}
Here, $k_{F_{N}}$ is the Fermi momentum, $E_{F_{N}}$ is the Fermi energy, and $\mu_{N}$ is the chemical potential of nucleon $N$. $m^*$ is the effective nucleon mass. The pressure $P$ is obtained using the Gibbs-Duhem relation:
\begin{equation} \label{eqn:pressure}
P = \sum_{N}{}\mu_{N}n_{N} - \epsilon~.
\end{equation}
We also add to this the contribution from leptons. In this work, the vector self-interaction ($\zeta$) term has been set to zero as it is known to soften the EoS \citep{mueller1996, tolos2017, pradhan2022zeta}.

In order to generate an EoS, we need to know the coupling constants ($g_{\sigma}$, $g_{\omega}$, $g_{\rho}$, $b$, $c$, $\Lambda_{\omega}$) in the Lagrangian. The isoscalar couplings ($g_{\sigma}$, $g_{\omega}$, $b$, $c$) are fitted using the nuclear saturation parameters ($n_{sat}$, $E_{sat}$, $K_{sat}$ and $m^*$) obtained from experiments. Similarly, the isovector couplings ($g_{\rho}$, $\Lambda_{\omega}$) are fitted to the symmetry properties ($E_{sym}$, $L_{sym}$) of nuclear matter at saturation \citep{hornick2018, Chen2014}. The empirical parameters in the pure nucleonic phase are therefore \{$n_{sat}$, $E_{sat}$, $K_{sat}$, $E_{sym}$, $L_{sym}$, $m^*$\}.

\subsubsection{Pure quark phase} \label{sec:quark}
For the pure quark phase, we use the MIT Bag model with first-order strong interaction corrections \citep{farhijaffe1984strange}. The grand potential density \citep{farhijaffe1984strange, Glendenning1997book} ($\Omega=-P$) is given by 

\begin{align}\label{GrandPotStrong}
\Omega=\sum_{f=u,d,s}\Omega_{f} + \sum_{l=e,\mu}\Omega_{l, free} + B~,
\end{align}

where $\Omega_{i, free}$ is the grand potential of free fermi gas and,

\begin{align}
    \Omega_{i} = \Omega_{i, free}&+\frac{1}{4\pi^2}\frac{2\alpha}{\pi}\Bigg[ \Bigg. 3\left(\mu_ik_{F_i}-m_i^2\ln{\frac{\mu_i+k_{F_i}}{\mu_i}}\right)^2  \nonumber \\
    &-  2k_{F_i}^2 - 3m_i^4\ln^2{\frac{m_i}{\mu_i}} \nonumber \\ 
    &+  6\ln{\frac{\tilde{\Lambda}}{\mu_i}}\left(m_i^2\mu_ik_{F_i} - m_i^4\ln{\frac{\mu_i+k_{F_i}}{m_i}}\right) \Bigg. \Bigg]~.
\end{align}

The energy density is calculated using Eqn. \ref{eqn:pressure}. 
$\alpha_s$ is the strong interaction coupling, and $\Tilde{\Lambda}$ is the QCD renormalization scale set to 300 MeV. We define $a_{4}=1-2\alpha_{s}/\pi$. The empirical parameters for the pure quark phase are the bag parameters \{$B^{1/4}$, $a_{4}$\}.

\subsubsection{Mixed phase: Gibbs construction} \label{sec:mixed}
In the quark deconfinement transition, two quantities are conserved - the baryon number and the total electric charge. Thus, we can characterize the phase transition by the Gibbs construction \citep{Glendenning1997book} where the hadronic matter and quark matter can coexist and is given by the condition
\begin{equation}\label{transition eqn}
    P_{H}(\mu_{b}, \mu_{q})=P_{Q}(\mu_{b}, \mu_{q})~.
\end{equation}
Here $P_{H}$ and $P_{Q}$ are the pressures of the hadron and quark phases, respectively, $\mu_{b}$ is the baryon chemical potential, and $\mu_{q}$ is the charge chemical potential. 
If we consider global charge conservation, we have the equations of constraint given by 
\begin{align}\label{globalchargeconservation}
\chi q^Q + (1-\chi) q^H&=0~, \nonumber \\
\chi n^Q + (1-\chi) n^H&=n_B~,
\end{align} 
where $Q$ stands for all quarks in the deconfined phase, $H$ stands for hadrons and leptons in the confined phase with $n$ and $q$ denoting the number density and electric charge, respectively. $n_B$ is the total baryon number density. The volume fraction ($\chi$) is defined as $\chi=V_{Q}/(V_{Q}+V_{H})$ for quarks, and the hadron volume fraction is given by $1-\chi$. The total energy density in the mixed phase is given by $$\epsilon=\chi\epsilon^{Q}+ (1-\chi)\epsilon^{H}~.$$

\subsection{Macroscopic structure} \label{sec:macro}
To calculate the mass ($M$) and radius ($R$) of non-rotating hybrid stars, we solve the Tolman-Oppenheimer-Volkoff (TOV) equations \citep{Glendenning1997book, schaffner-bielich_2020}

\begin{align}\label{TOV equations}
    \frac{dP(r)}{dr} &= -\frac{[P(r)+\epsilon(r)][m(r)+4\pi r^3P(r)]}{r(r-2m(r))}~, \nonumber \\
    \frac{dm(r)}{dr}&=4\pi r^{2}\epsilon(r)~,
\end{align}

along with the equation state, applying the boundary conditions $m(r=0)=0$ and $P(r=R)=0$. The dimensionless tidal deformability ($\Lambda$) is defined as \begin{equation}\label{Lambda}
\Lambda=\frac{2}{3}\frac{k_{2}}{C^{5}}~,
\end{equation}

where $C=GM/Rc^{2}$ is the dimensionless compactness. To calculate this we solve for the $l = 2$ love number $k_{2}$ as done in
\citet{flanagan2008ads, hinderer2008ads, damour2009, yagi2013prd}.

\subsection{Cut-off filter scheme} \label{sec:bayesian}
In this study, we apply a ``cut-off filter" scheme where we impose strict limits demanding compatibility with nuclear and astrophysical observational data to obtain the posteriors. As in Bayesian analysis, the priors are obtained by varying the empirical nuclear and quark model parameters described in Sec.~\ref{sec:prior}, and the likelihood functions are appropriately chosen physical conditions as filter functions as explained in Sec.~\ref{sec:constraints}. We then use the posterior sets to calculate correlations (Sec.~\ref{sec:correlations}). In our previous work~\citep{ghosh2022multi}, it was explicitly demonstrated that including the statistical re-weighting using $\chi$-squared statistics might change the posterior probability distribution slightly, but it does not significantly alter the physical correlation between nuclear empirical parameters and astrophysical observables. Many other works have followed the same scheme \citep{kurkela2014constraining, Most2018radiuslamfromgw, annala2018gwconstraints, annala2020evidence, ghosh2022multihyperon}. A similar scheme was employed to study the correlations between nuclear parameters in various works \citep{chatterjee2017, magueron2018, ghosh2022multi, ghosh2022multihyperon}. We adopt this ``cut-off filter" scheme for this investigation. 

\subsubsection{Priors}\label{sec:prior}
As explained in sections \ref{sec:nucleonic} and \ref{sec:quark}, the combined set of empirical parameters is \{$n_{sat}$, $E_{sat}$, $K_{sat}$, $E_{sym}$, $L_{sym}$, $m^*$, $B^{1/4}$, $a_{4}$\}. We vary these model parameters uniformly (flat prior) within a prior range given in Table \ref{table:priorranges} as was done in our previous work~\citep{ghosh2022multi}

\begin{deluxetable*}{cccccccc}\label{table:priorranges} 
\tablecaption{The range of variation of empirical nuclear and Bag model parameters within their present uncertainties.}
\tablehead{\colhead{$n_{sat}$} & \colhead{$E_{sat}$} & \colhead{$K_{sat}$} & \colhead{$E_{sym}$} & \colhead{$L_{sym}$} & \colhead{$ m^*/m$} & \colhead{$a_4$} & \colhead{$B^{1/4}$} \\
\colhead{(fm$^{-3}$)} & \colhead{(MeV)} & \colhead{(MeV)} & \colhead{(MeV)} & \colhead{(MeV)} & & & \colhead{(MeV)}} 
\startdata
0.14 - 0.17  & -16.0 $\pm$ 0.2 & 200 - 300 & 28 - 34   & 40 - 70 & 0.55 - 0.75 & 0.4 - 1.0 & 100 - 300
\enddata
\tablecomments{Meson masses are set to $m_{\sigma}=550$ MeV, $m_{\omega}=783$ MeV, $m_{\rho}=770$ MeV, the quark masses to $m_u=5$ MeV, $m_d=5$ MeV, $m_s=100$ MeV and the nucleon mass is set to $m_N=939$ MeV.}
\end{deluxetable*}

We generate EoSs with model parameters randomly varied in the range shown in Table~\ref{table:priorranges}. The range of variation is as used in our previous work \citep{ghosh2022multi} given by state-of-the-art nuclear experimental data \citep{oertel2017, hornick2018}. The range for parameter $a_4$ is chosen so that at $a_4=1$, $\alpha_s=0$ and for $a_4=1$, $\alpha_s\approx1$ beyond which perturbation theory does not apply. A wide range of values is considered for $B^{1/4}$ (100-300 MeV). The lower limit is set to $B^{1/4}=100$ MeV, as for lower values, the mixed phase begins within the crust \citep{nandi2018}. For large values of the Bag constant, the hybrid EoS asymptotically resembles the pure nucleonic EoS, as the mixed phase is not reached for the densities attained in the NS interior. We set an upper limit of $B^{1/4}=300$ MeV as done in \citet{weissenborn2011quark}.

\subsubsection{Constraints}\label{sec:constraints}
Firstly, for an EoS generated for a set of model parameters, we check if it is physical by verifying that it satisfies the physical conditions such as positive-definiteness of pressure, energy, speed of sound, and causality ($c_s^2 < 1$).  We then apply various cut-off filters to constrain the model parameter space. The following multi-physics constraints are used in this work:
\begin{enumerate}
    \item \textbf{CEFT at low densities}\\
    The EoS at low densities (0.07-0.20 fm$^{-3}$) is constrained using recent calculations from CEFT \citep{drischler2019}. CEFT takes into account the many-body interactions and solves for the nuclear forces in the low momentum limit. These microscopic calculations are only valid for pure neutron matter (PNM). We construct the PNM EoS using the nuclear saturation parameters from Table~\ref{table:priorranges} and compare it with the CEFT calculations. If the EoS falls in the CEFT uncertainty band, we retain this set of parameters which is then used to construct hybrid EoS for varying bag parameters. These hybrid EoSs are further checked for consistency with other constraints explained below.
    \item \textbf{Astrophysical observations at high densities}\\
    In this study, we use the following state-of-the-art multi-messenger (electromagnetic and gravitational-wave) astrophysical observations:
    \begin{enumerate}
        \item  A recent observation of the high-mass pulsar, PSR J0740+6620, estimates its mass to be $2.08^{+0.07}_{-0.07}M_{\odot}$ \citep{Fonseca_2021}, the errors indicating $1\sigma$ interval. Thus, we use $2.01M_{\odot}$ as the lower limit on the maximum mass of NS EoS. Only the EoSs that allow for NS masses larger than this limit are allowed.
        \item The GW signal from the binary NS merger event GW170817 \citep{Abbott2017AGW170817, abbott2017BGW170817multi} has been analyzed to obtain various properties of the constituent NSs \citep{abbottetal2019properties}. One such crucial entity is the 90\% highest posterior density interval for tidal deformability in the case of low-spin prior measured as $\Tilde{\Lambda}=300^{+420}_{-230}$ \citep{abbottetal2019properties}. 
       This leads to an upper limit for the estimate for the tidal deformability of a 1.4$M_{\odot}$ NS as $\Lambda_{1.4M_{\odot}}\leq720$ \citep{abbottetal2019properties, TongZhao2020R1.4}. For any given EoS, we calculate $\Lambda$ for the mass of $1.4M_{\odot}$, and this value should fall below 720 in order to say that it satisfies this constraint.
    \end{enumerate}
    \item \textbf{pQCD at very high densities}\\
    The EoS of cold, dense matter can be calculated at very high densities using perturbative techniques. This is done up to second order in strong coupling constant ($\mathcal{O}(\alpha_{s}^2)$) \citep{kurkela2010pqcd}. The pressure as a function of chemical potential can be evaluated using a fitting function for cold quark matter \citep{fraga2014pqcd}, which depends on the renormalization scale $\Tilde{\Lambda}$. It is given by the following function:
\begin{equation}
    P(\mu_{B})=P_{free}(\mu_{B})\left(c_{1}-\frac{a(X)}{(\mu_{B}/GeV)-b(X)}\right)~,
\end{equation}
\begin{equation}
    a(X) = d_{1}X^{-\nu_{1}},\; b(X) = d_{2}X^{-\nu_{2}}~,
\end{equation}
where $P_{free}$ is the pressure of the three non-interacting massless quarks:
\begin{equation}
    P_{free}(\mu_{B}) = \frac{3}{4\pi^2}\left(\frac{\mu_{B}}{3}\right)^4~.
\end{equation}
$X$ is a dimensionless parameter that depends on the renormalization scale as $X\equiv3\Tilde{\Lambda}/\mu_{B}$, which belongs to the range $X \in [1,4]$ for a good fit. The best fit parameters are given by \citet{fraga2014pqcd}:
\begin{equation}
    c_{1}=0.9008, d_{1}=0.5034, d_{2}=1.452, \nonumber
\end{equation}
\begin{equation}
    \nu_{1}=0.3553; \nu_{2}=0.9101~.
\end{equation}
For a set of empirical parameters, we generate the EoS up to high densities. For
baryon chemical potentials $\mu_B=2.6$ GeV (corresponding to baryonic number densities $n_B \ge 40 n_0$), we verify the pressure of the hybrid EoS falls within the pQCD range, i.e., matches the pQCD pressure for some value of $X \in [1,4]$.
\end{enumerate}

\subsubsection{Correlations}\label{sec:correlations}
After applying the filter functions mentioned in the previous section, only the EoSs satisfying the constraints survive, which we call the posterior set. Using this posterior, we can study the correlations between the physical parameters of the model amongst themselves and with the NS observables like mass, radius, and tidal deformability. We evaluate the Pearson's correlation coefficient between a pair of quantities $X, Y$ as
$$ r_{XY} = \frac{Cov(X,Y)}{\sqrt{Cov(X,X)}\sqrt{Cov(Y,Y)}}~.$$
$Cov(X, Y)$ is the covariance between the quantities $X$ and $Y$ defined as
$$Cov(X,Y)=\frac{1}{N}\sum_{i=1}^{N}(X_i-\bar{X})(Y_i-\Bar{Y})~,$$
where $\Bar{X}$ is the sample mean given by $\Bar{X}=\frac{1}{N}\sum_iX_i~.$

\section{Results}\label{sec:results}
For each EoS, we generate the corresponding pure neutron matter (PNM) EoS using only the nuclear parameters, and the posterior set consistent with CEFT and other physical conditions such as positive-definiteness of pressure, energy, speed of sound and causality ($c_s^2<1$) is referred to as ``CEFT". Out of the 100,000 parameter sets, about 23$\%$ satisfy these constraints. The posterior set is then passed through astrophysical constraint filters outlined in Sec.~\ref{sec:constraints} and referred to as ``Astro". This scheme allows 53$\%$ of the EoSs from the CEFT posterior set. If we only add the pQCD constraint instead of astrophysical ones, we are left with 52$\%$. Finally, both the astrophysical and pQCD constraints are applied, permitting approximately 28$\%$ EoSs through both filters.

\subsection{Posteriors}

For the posterior sets obtained after passing through the different constraint filters, we analyze the range of densities and baryon chemical potentials in the NS core. For the maximum mass configuration for each EoS, the minimum and maximum values of the central density ($n_{cen}$) and the corresponding baryon chemical potential ($\mu_{cen}$) are provided in Table.~\ref{table:densityminmax}. The values are compatible with those in \citet{kurkela2014constraining}, where a maximum central density of 8.0 $n_{sat}$ was obtained, although we find an upper bound on $\mu_{cen}$ that is higher by 0.2 GeV. We also note down the range of maximum mass ($M_{max}$), radius ($R_{1.4M_{\odot}}$) and tidal deformability ($\Lambda_{1.4M_{\odot}}$) from the posterior set in Table.~\ref{table:densityminmax}.

\begin{deluxetable*}{cccccc}\label{table:densityminmax} 
\tablecaption{The range of values observed after application of different filters.}
\tablehead{\colhead{Posterior} & \colhead{$n_{cen}$} & \colhead{$\mu_{cen}$} & \colhead{$M_{max}$}  & \colhead{$R_{1.4M_{\odot}}$} & \colhead{$\Lambda_{1.4M_{\odot}}$}\\
&\colhead{(fm$^{-3}$)}&\colhead{(GeV)}&\colhead{($M_{\odot}$)}&\colhead{(km)}&} 
\startdata
CEFT & 0.54 - 1.77 &  1.13 - 1.98 & 1.04 - 2.99 & 8.8 - 14.2 & 33 - 1263\\ 
CEFT+Astro & 0.68 - 1.23 & 1.33 - 1.98 & 2.01 - 2.67 & 11.3 - 13.5 & 247 - 720\\ 
CEFT+Astro+pQCD & 0.69 - 1.23 & 1.33 - 1.97 & 2.01 - 2.67 & 11.4 - 13.5 & 247 - 720\\
\enddata
\tablecomments{The range for central density ($n_{cen}$) and chemical potential ($\mu_{cen}$) reached inside a maximally massive NS, maximum TOV mass ($M_{max}$), radius ($R_{1.4M_{\odot}}$) and tidal deformability ($\Lambda_{1.4M_{\odot}}$) of a 1.4 $M_{\odot}$  for each posterior set have been mentioned. The first row corresponds to the CEFT posterior set containing EoSs that satisfy CEFT constraints and physical requirements. The second row corresponds to CEFT+Astro set, where we add the astrophysical constraints ($M \ge 2.01 M_{\odot}$ and $\Lambda_{1.4M_{\odot}} \le 720$). The last row corresponds to CEFT+Astro+pQCD set of EoSs that additionally satisfies constraint from pQCD at high densities.}
\end{deluxetable*}

\begin{figure}
    \epsscale{1.15}
    \plotone{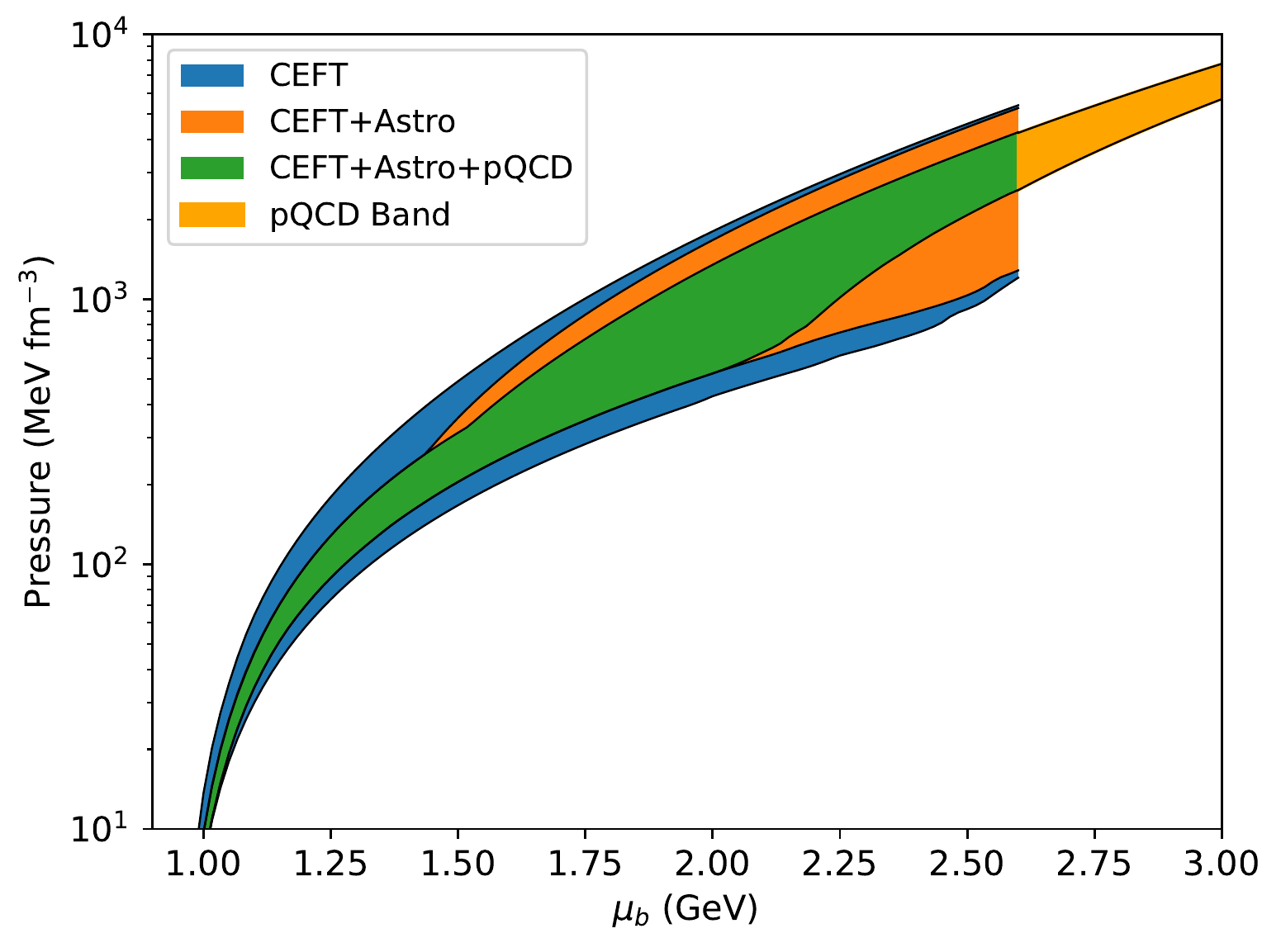}
    \caption{EoS (pressure vs baryon chemical potential) contours obtained on imposing filters from Sec.~\ref{sec:constraints}. The blue contour is obtained after enforcing only CEFT and minimal physics constraints, the orange contour on further imposing astrophysical constraints, while the green contour when we additionally impose pQCD constraint}
    \label{fig:EoS}
\end{figure}

\begin{figure}
    \epsscale{1.15}
    \plotone{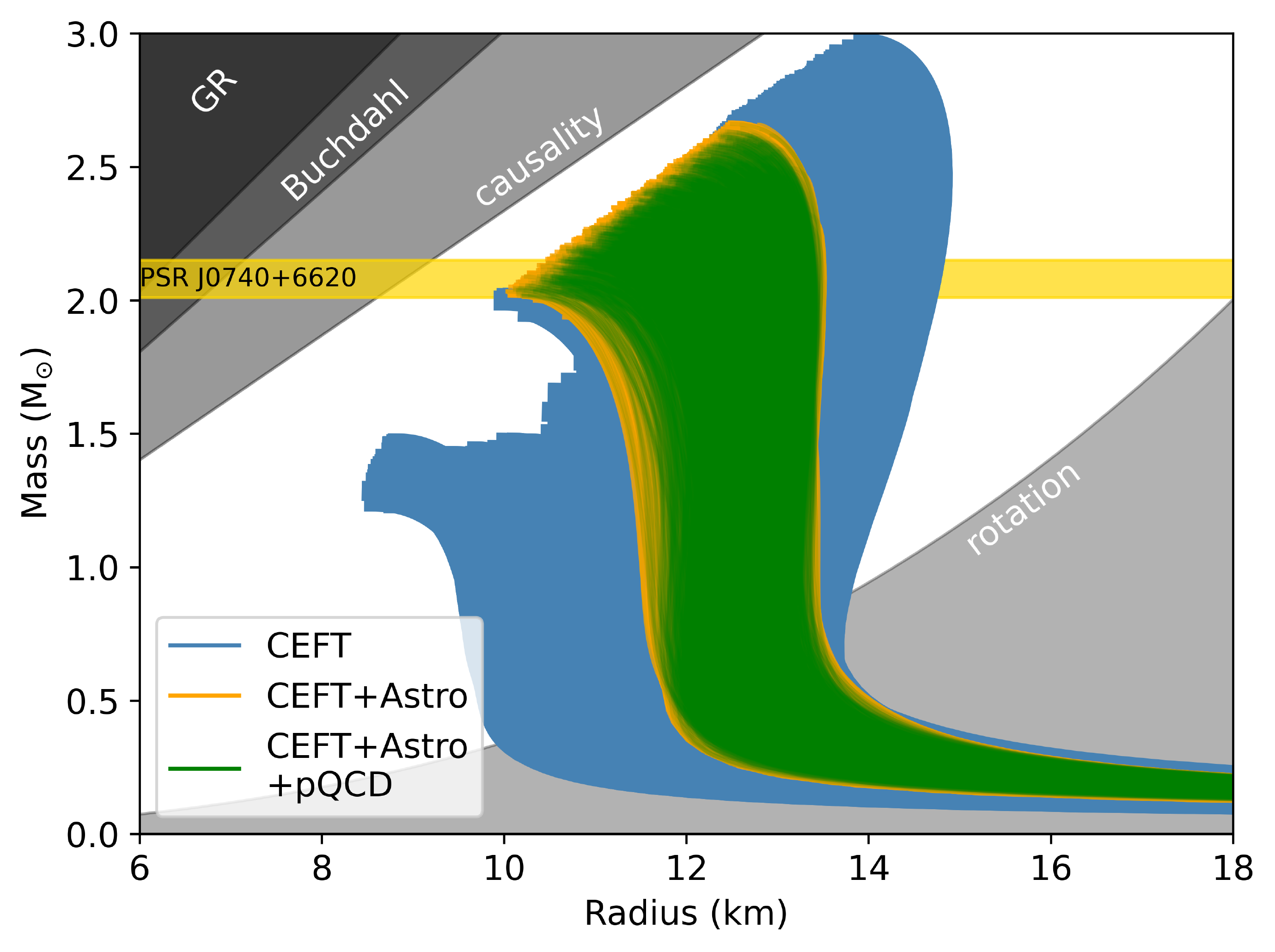}
    \caption{Mass-Radius contours corresponding to the EoS contours shown in Fig. \ref{fig:EoS}}
    \label{fig:MR}
\end{figure}

\begin{figure}
    \epsscale{1.15}
    \plotone{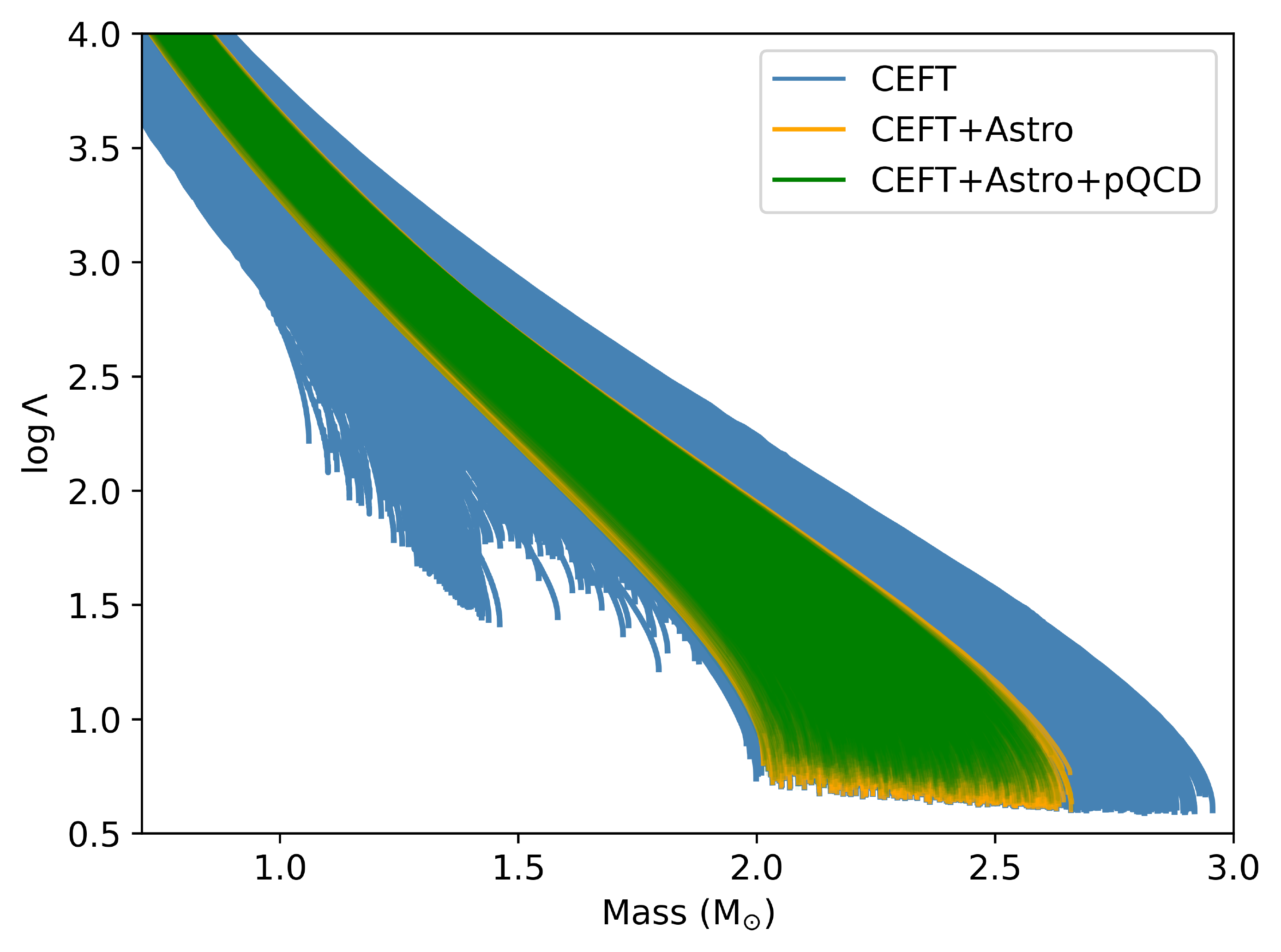}
    \caption{Dimensionless Tidal Deformability-Mass contours corresponding to the EoS contours shown in Fig. \ref{fig:EoS}}
    \label{fig:LM}
\end{figure}

In the following section, we display the posteriors obtained after passing the prior sample through the different constraint filters. In Fig. \ref{fig:EoS}, the EoS is represented by the pressure plotted as a function of baryon chemical potential. A broad blue band is obtained for EoS with the imposition of CEFT alone. As CEFT constrains the EoS only at low densities, the band broadens at higher chemical potentials. The addition of astrophysical constraints narrows the band further (orange), particularly at intermediate densities up to $\mu_B=1.5$ GeV, beyond which the band broadens again. At higher chemical potentials, the solutions obtained are unstable. Although pQCD constricts the band at higher densities (green) beyond  $\sim \mu_B=2.0$ GeV, such densities are not reached in the NS interior (refer Table. \ref{table:densityminmax}). Below $\sim \mu_B=1.5$ GeV, the pQCD cloud coincides with the orange cloud, and it is not very constraining below $\mu_B=2.0$ GeV. This can be verified from Figs. \ref{fig:MR} and \ref{fig:LM} where we plot the mass vs. radius and tidal deformability vs. mass contours, respectively, for the EoS contours corresponding to Fig.~\ref{fig:EoS}. We can observe that the addition of pQCD constraints only weakly affects the NS observables, i.e., the orange and green bands nearly coincide. It is well known that pQCD is most instrumental when constraining the higher density regime of the EoS. However, the pQCD constraint does improve our understanding of the model parameters, as we will discuss later.

\newpage
\newpage
\subsection{Correlation Study}\label{sec:res_bayesian}
\begin{figure}
    \epsscale{1.15}
    \plotone{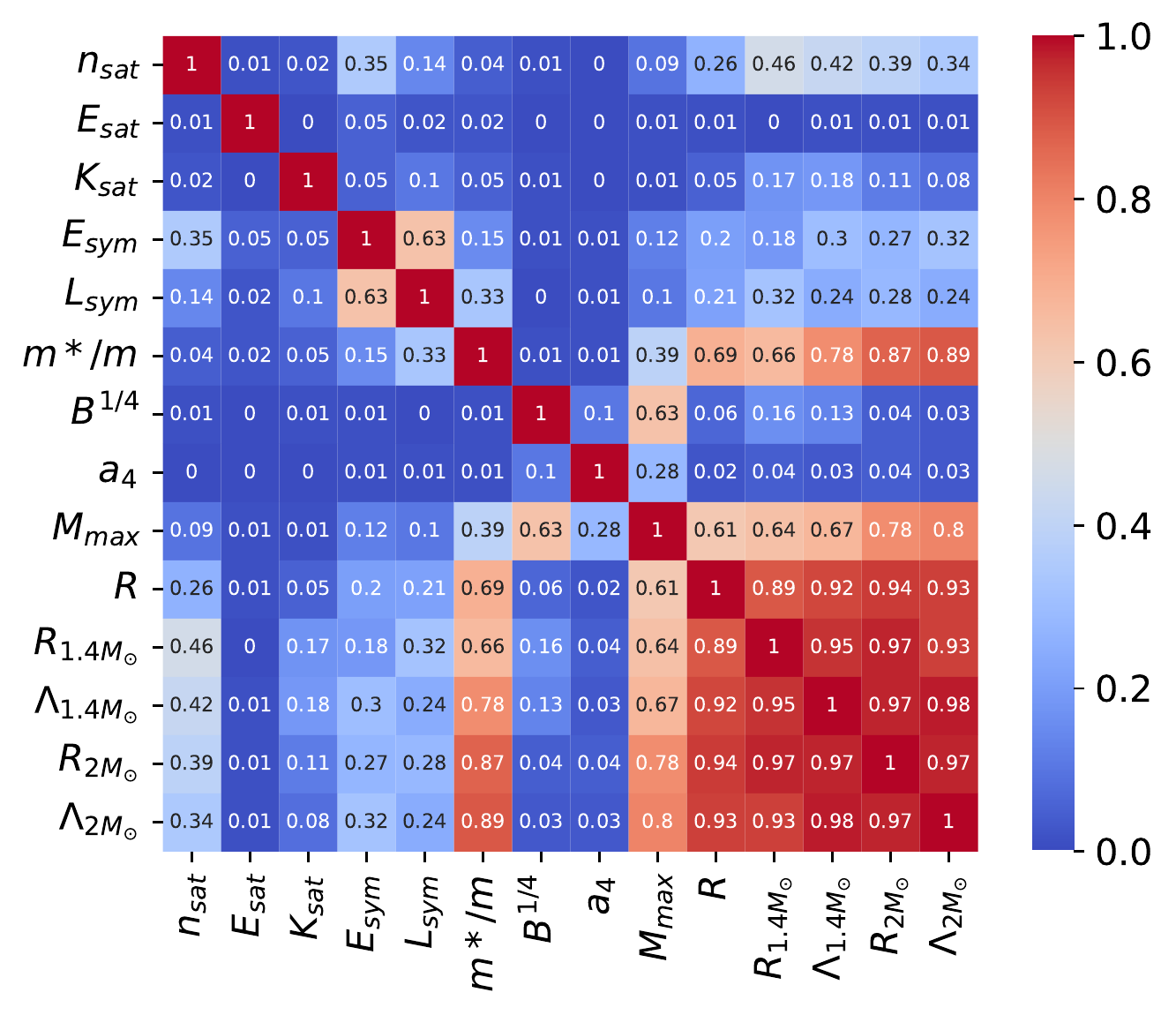}
    \caption{Correlation coefficients between the nuclear and quark model parameters along with the NS observables. These are after applying the CEFT and minimal physics filters}
    \label{fig:correlation_ceft}
\end{figure}

Using the posterior sets discussed in the previous section, we investigation possible correlations among the  empirical parameters of both the nuclear ($n_{sat}$, $E_{sat}$, $K_{sat}$, $E_{sym}$, $L_{sym}$, $m^*/m$), and the quark model ($B^{1/4}$, $a_4$), as well as with the NS observables ($M_{max}$, $R$, $R_{1.4 M_{\odot}}$, $R_{2 M_{\odot}}$, $\Lambda_{1.4 M_{\odot}}$, $\Lambda_{2 M_{\odot}}$), where $R$ is the radius corresponding to $M_{max}$. From a plot of the correlation coefficient matrix in Fig. \ref{fig:correlation_ceft} after applying CEFT constraints, we draw the following conclusions:
\begin{itemize}
    \item All NS observables are strongly correlated with each other ($>0.9$), although the correlations of $M_{max}$ are relatively smaller (0.6-0.8).
    \item $n_{sat}$ is moderately correlated with the NS observables of 1.4 $M_{\odot}$ and 2 $M_{\odot}$ ($\sim$0.4). The correlation with $R_{1.4 M_{\odot}}$ is noticeable (0.46), that with $R$ is weak (0.26) and that with $M_{max}$ is negligible. There is a weak correlation between $L_{sym}$ and $R_{1.4 M_{\odot}}$ (0.32). $m^*/m$ is strongly correlated with NS observables (0.7-0.9) but is only moderately correlated with $M_{max}$ (0.4).
    \item $E_{sym}$ and $L_{sym}$ show moderate correlation (0.63).
    \item $M_{max}$ is moderately correlated with $B^{1/4}$ (0.63) and weakly with $a_4$ (0.28). We do not find a correlation of $B^{1/4}$ and $a_4$ with any other NS observable or empirical nuclear parameter.
\end{itemize}

\begin{figure}
    \epsscale{1.15}
    \plotone{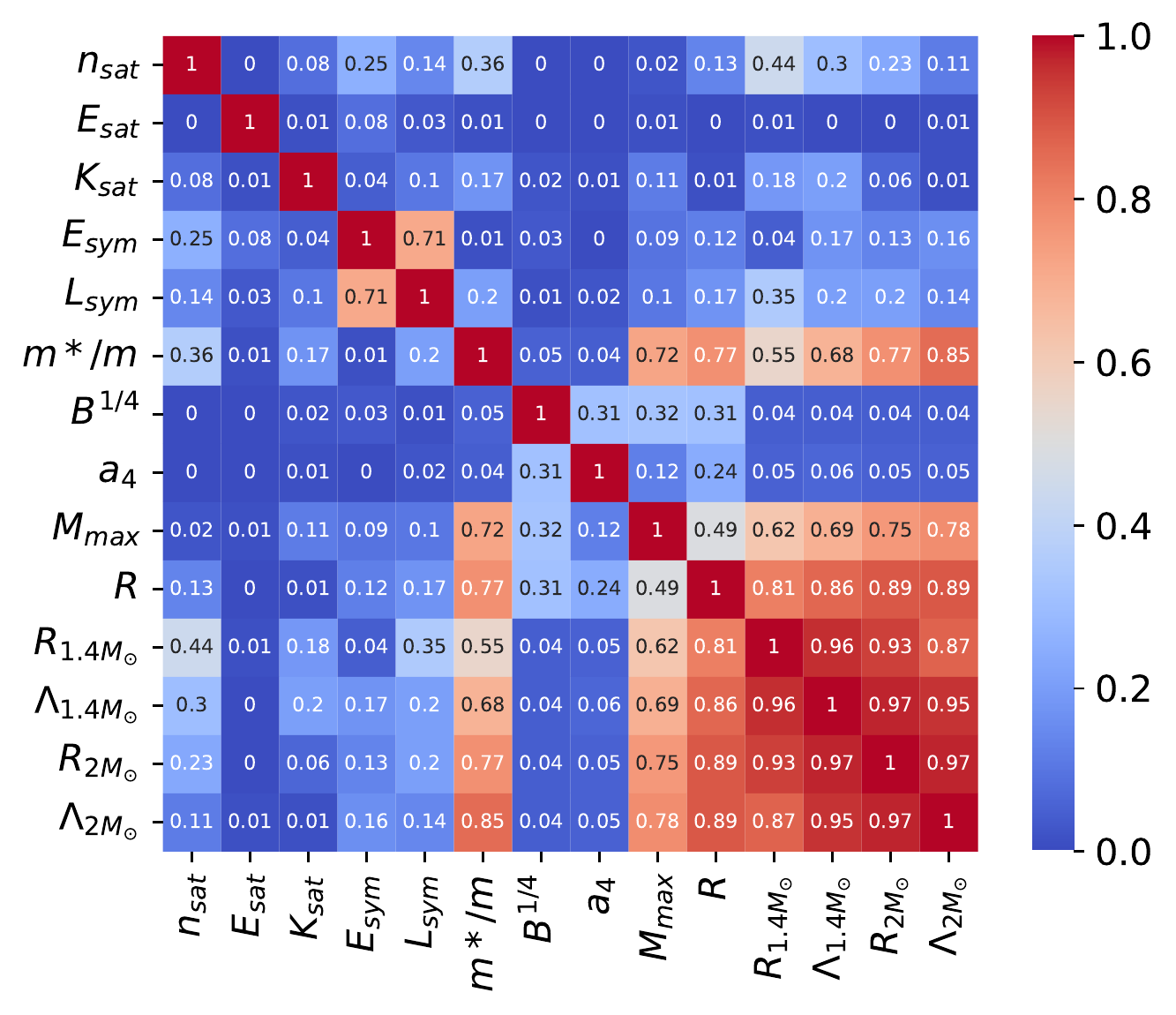}
    \caption{Posterior correlation matrix of the nuclear and quark model parameters along with the NS observables after the application of CEFT and astrophysical constraints filters as explained in Sec. \ref{sec:constraints}}
    \label{fig:correlation_ceft_astro}
\end{figure}

We then add the astrophysical constraint filters as explained in Sec. \ref{sec:constraints}. We plot the correlation matrix from this posterior in Fig. \ref{fig:correlation_ceft_astro} and observe the following changes:
\begin{itemize}
    \item The correlation between $M_{max}$ and $R$ reduces from 0.61 to 0.49. Those between the rest of the NS observables stay unaltered.
    \item The correlations of $n_{sat}$ with NS observables become weak ($<0.3$) except for that with $R_{1.4 M_{\odot}}$ (0.44). The correlation between $L_{sym}$ and $R_{1.4 M_{\odot}}$ remains unaffected (0.35). The correlation between $m^*/m$ and $M_{max}$ becomes strong as it increases from 0.4 to 0.72, making $m^*/m$ the most important parameter in determining NS observables.
    \item A weak correlation between $n_{sat}$ and $m^*/m$ appears (0.36). $E_{sym}$ and $L_{sym}$ now show a strong correlation (0.71).
    \item The correlation of $M_{max}$ with $B^{1/4}$ becomes weak (0.32) and that with $a_4$ becomes negligible (0.12). Weak correlations appear for $R$ with $B^{1/4}$ (0.31) and $a_4$ (0.24). $B^{1/4}$ and $a_4$ become weakly correlated (0.31).
\end{itemize}

Similar conclusions about the correlations between nuclear and astrophysical observables were drawn in\citet{ghosh2022multi}. For the pure nucleonic EoS, $L_{sym}$ and $E_{sym}$ showed strong correlation and a moderate correlation between $L_{sym}$ and $m^*/m$ (see Fig. 5 in~\citet{ghosh2022multi}) was observed which was due to the CEFT filter. For astrophysical observables, $n_{sat}$ and $m^*/m$ had significant correlations although the correlation with $n_{sat}$ was much weaker for 2M$_{\odot}$ stars. Here also for neutron stars with a hybrid EoS, we observe similar correlations between the nuclear parameters and astrophysical observables. The correlation between $R$ and $\Lambda$ of 1.4 and 2 $M_{\odot}$ stars is expected from Eq.~\ref{Lambda}. Several studies reported a correlation between $L_{sym}$ and $R_{1.4 M_{\odot}}$ \citep{fattoyev2013, alam2016, lim2018, zhu2018}. \citet{zhang2019} also looked for a relation between $L_{sym}$, $R_{1.4 M_{\odot}}$ and $\Lambda_{1.4 M_{\odot}}$ However, our finding is consistent with more recent studies which find them to be nearly independent \citep{hornick2018, ghosh2022multi, ghosh2022multihyperon}. A recent study by \citet{biswas2021} also reported a weak correlation between $L_{sym}$ and $R_{1.4 M_{\odot}}$, where laboratory experiments and astrophysical observations, including NICER observations, were used in a Bayesian framework.
Our results are also in agreement with \citet{hornick2018}, where it was concluded that $m^*/m$ is more important in determining $R_{1.4 M_{\odot}}$ than $L_{sym}$ when CEFT effects are considered. Lower values of $m^*/m$ stiffen the EoS and, thus, raises the maximum mass and radius. This follows from the Hugenholtz-van-Hove theorem explained in \citet{hornick2018}. When quark matter is included in the model, we find that $M_{max}$ is controlled more by $B^{1/4}$ than $m^*/m$ when only CEFT effects are included (See Fig.~\ref{fig:correlation_ceft}). However, inclusion of astrophysical constraints makes $m^*/m$ the most dominant parameter for determining NS observables (See Fig.~\ref{fig:correlation_ceft_astro}).

\begin{figure}
    \epsscale{1.15}
    \plotone{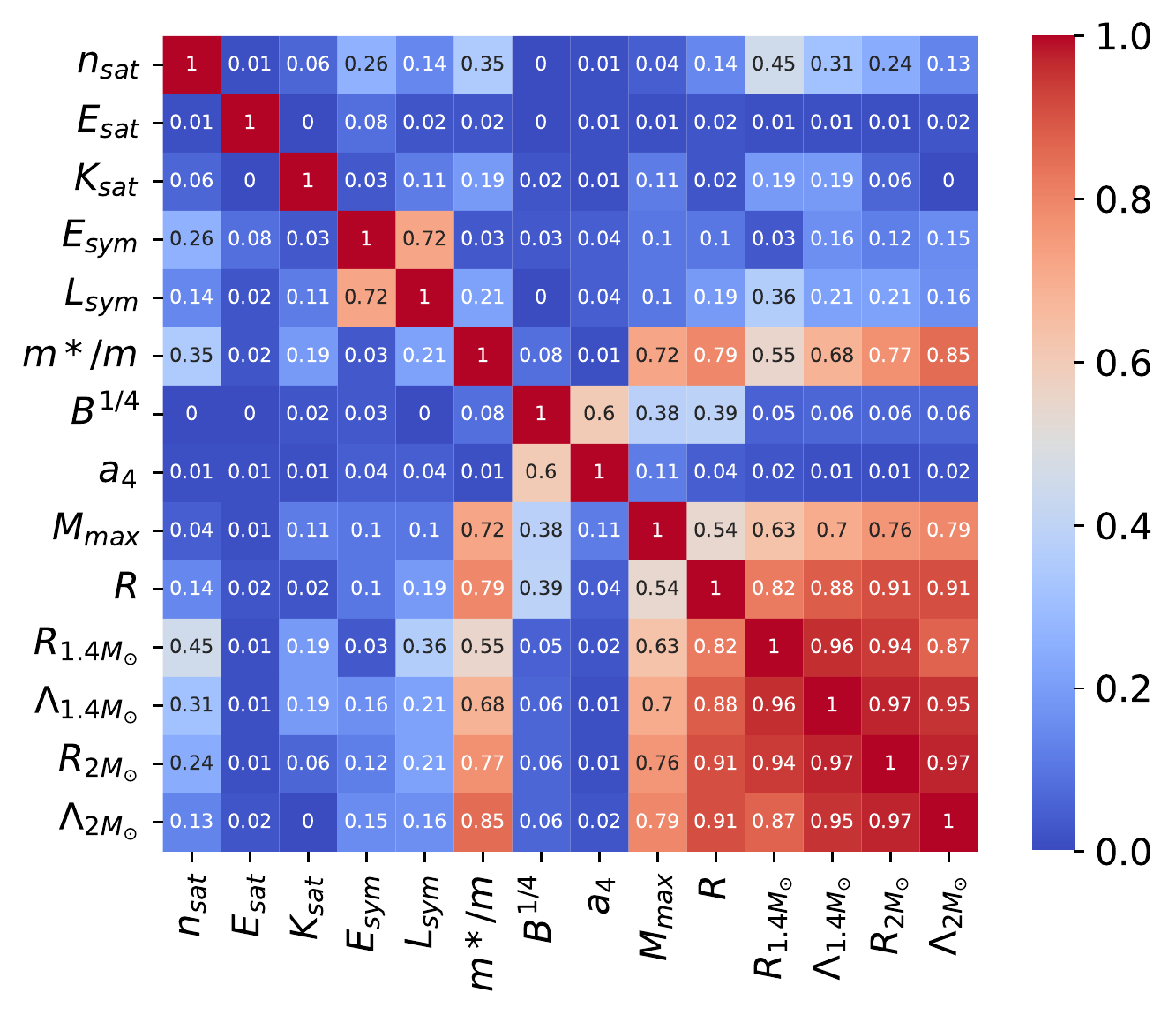}
    \caption{The correlation matrix same as in Fig. \ref{fig:correlation_ceft_astro} after imposing the pQCD filter as explained in Sec. \ref{sec:constraints}}
    \label{fig:correlation_full}
\end{figure}

Finally, in Fig. \ref{fig:correlation_full}, we plot the correlation matrix after imposing the pQCD constraint additionally and conclude the following:
\begin{itemize}
    \item The correlations of nuclear empirical parameters and NS observables remain unaffected.
    \item The correlations of $B^{1/4}$ with $M_{max}$ and $R$ increase to $\sim$0.4. Correlation of $a_4$ with $R$ becomes negligible. We do not find a correlation of $a_4$ with any parameter, except for a moderate correlation (0.6) with $B^{1/4}$. 
\end{itemize}

\begin{figure*}
    \epsscale{1.15}
    \plotone{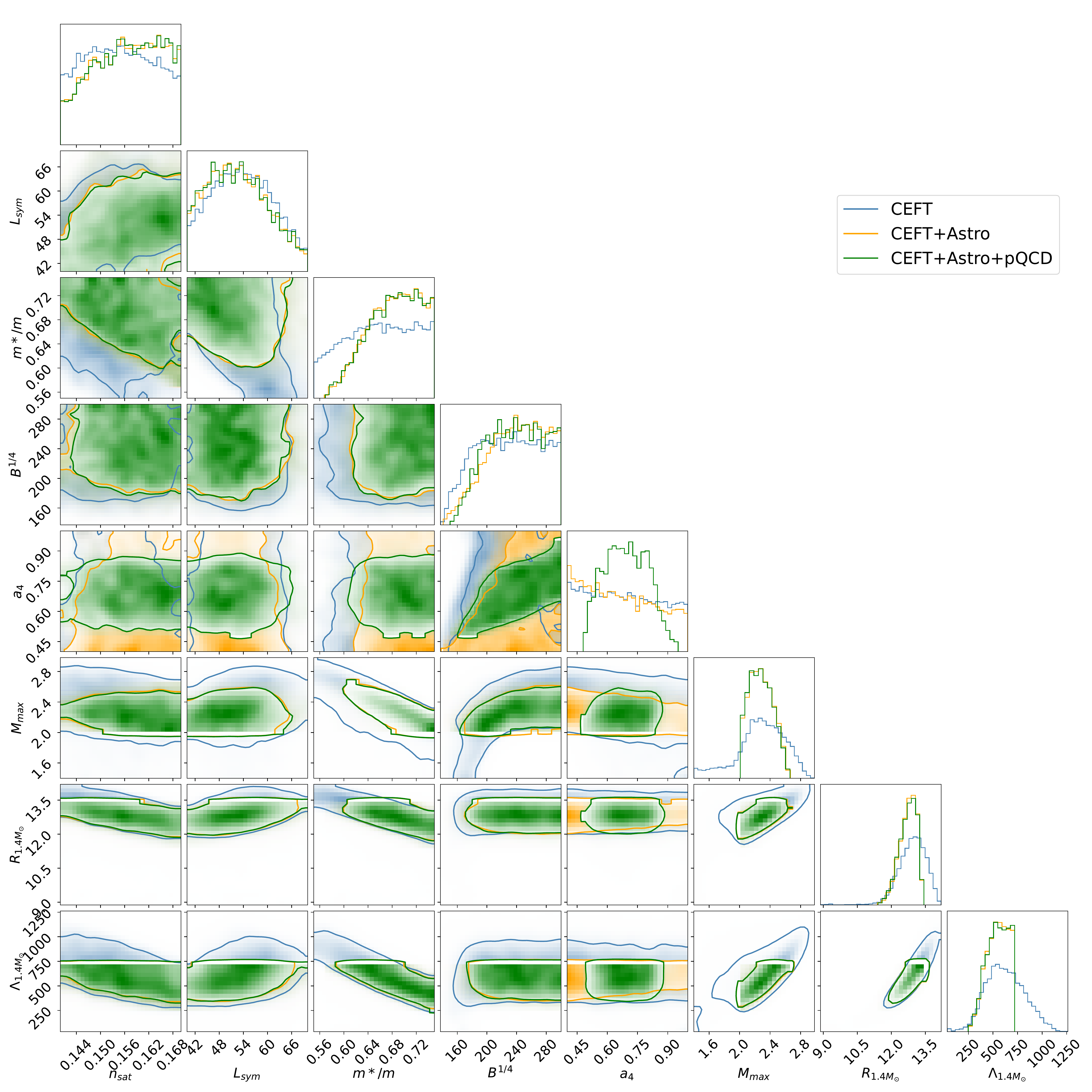}
    \caption{Posterior distributions of model parameters and NS observables that are strongly correlated presented in a corner plot after application of various filters. The colour scheme is same as in Fig. \ref{fig:EoS}.}
    \label{fig:corner}
\end{figure*}

\begin{figure}
    \epsscale{1.15}
    \plotone{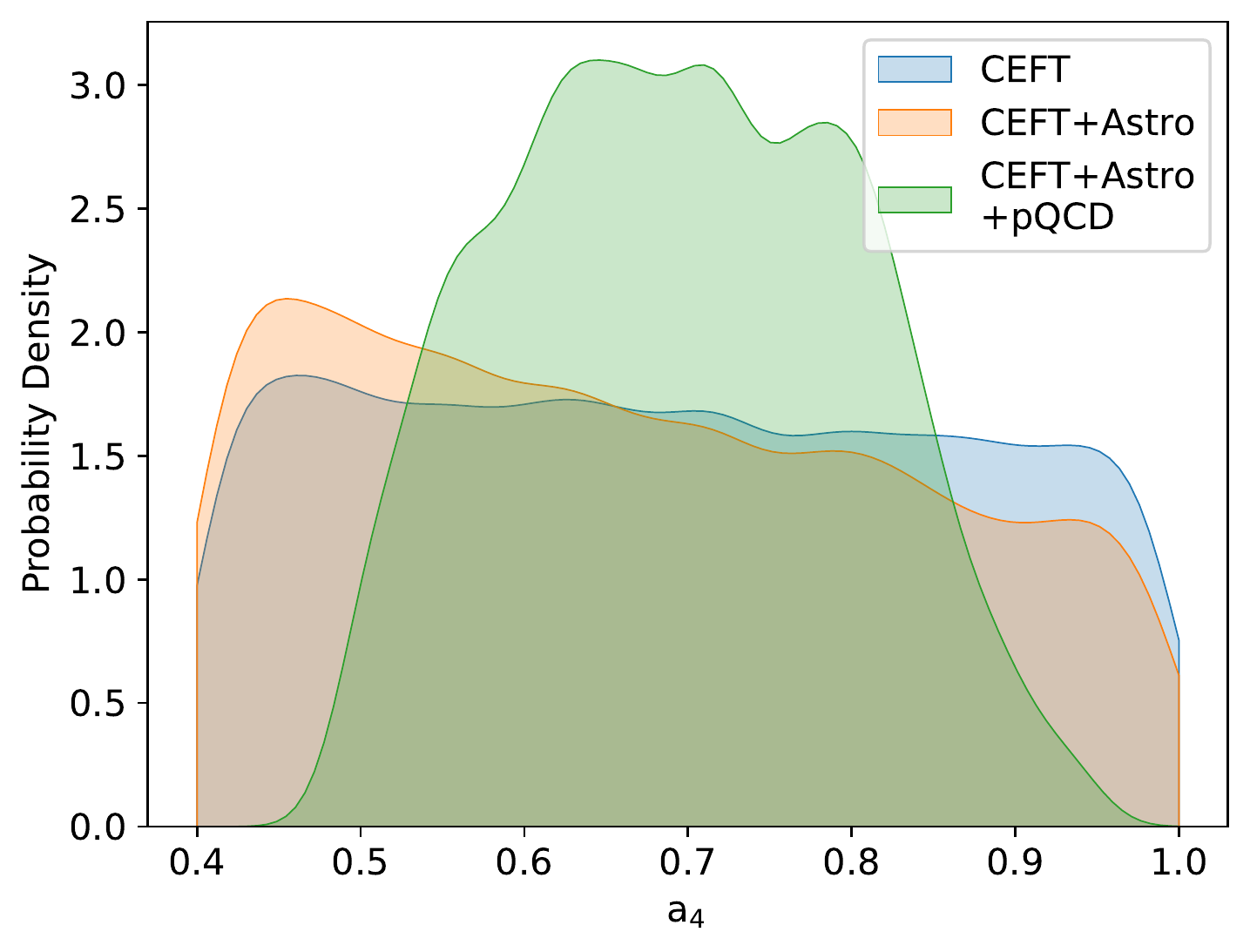}
    \caption{The change in the probability distribution of bag model parameter $a_4$ as a result of different constraint filters}
    \label{fig:a4}
\end{figure}

For a better understanding, we show the posterior distributions of empirical parameters  ($n_{sat}$, $L_{sym}$, $m^*/m$, $B^{1/4}$ and $a_4$) and NS observables in a corner plot (Fig. \ref{fig:corner}). We observe from the corner plot that lower values of $L_{sym}$ and $m^*/m$ are disfavoured (compared with a flat prior) just after imposing CEFT constraints. This is consistent with the findings of \citet{hornick2018, ghosh2022multi} as low values lead to unphysical EoSs. \citet{hornick2018} also showed that for values of $L_{sym}$ larger than about 60, the EoS falls outside the CEFT band. Values of $L_{sym}\gtrsim60$ MeV are incompatible with combined constraints from terrestrial experiments, astrophysical observations, and theoretical calculations \citep{lattimer2013} although PREX II experiment implies a much higher value of $L_{sym} = 106 \pm 37 MeV$~\citep{EssickPrex2021} from the measurement neutron skin thickness of $^{208}$Pb~\citep{Prex2021}. Adding quark degrees of freedom softens EoS and lowers the mass of NS. We can expect $m^*/m$ to reduce to compensate for this effect. Within the uncertainty range of the empirical parameters of this study, however, we can have stiff EoSs without requiring a low value of $m^*/m$. We thus get a similar posterior for the effective nucleon mass as in pure nucleonic case \citep{ghosh2022multi} peaking around 0.7. The excessive softening of the EoS is countered by a delayed onset of the hadron-quark phase transition achieved mainly from higher values of $B^{1/4}$ and lower values of $a_4$ \citep{weissenborn2011quark}. Thus, the distribution of $B^{1/4}$ shifts to higher values, and that of $a_4$ shifts slightly to lower values (also see Fig.~\ref{fig:a4}) after adding astrophysical constraints. We observe that the distribution is not flat for the CEFT case of $B^{1/4}$. This is because low values of $B^{1/4}$ are unphysical, where the mixed phase begins within the crust. Particularly interesting is the posterior distribution of the $a_4$ parameter, which becomes peaked only after imposing the pQCD constraint. The normalized, smoothened posterior distributions of $a_4$ after the application of various filters are highlighted in Fig. \ref{fig:a4}. EoS falls outside the pQCD band for extreme values of $a_4$, and the distribution peaks roughly in the central area (0.55-0.85).

\newpage
\subsection{Implications for the quark EoS}

\begin{figure}
    \epsscale{1.15}
    \plotone{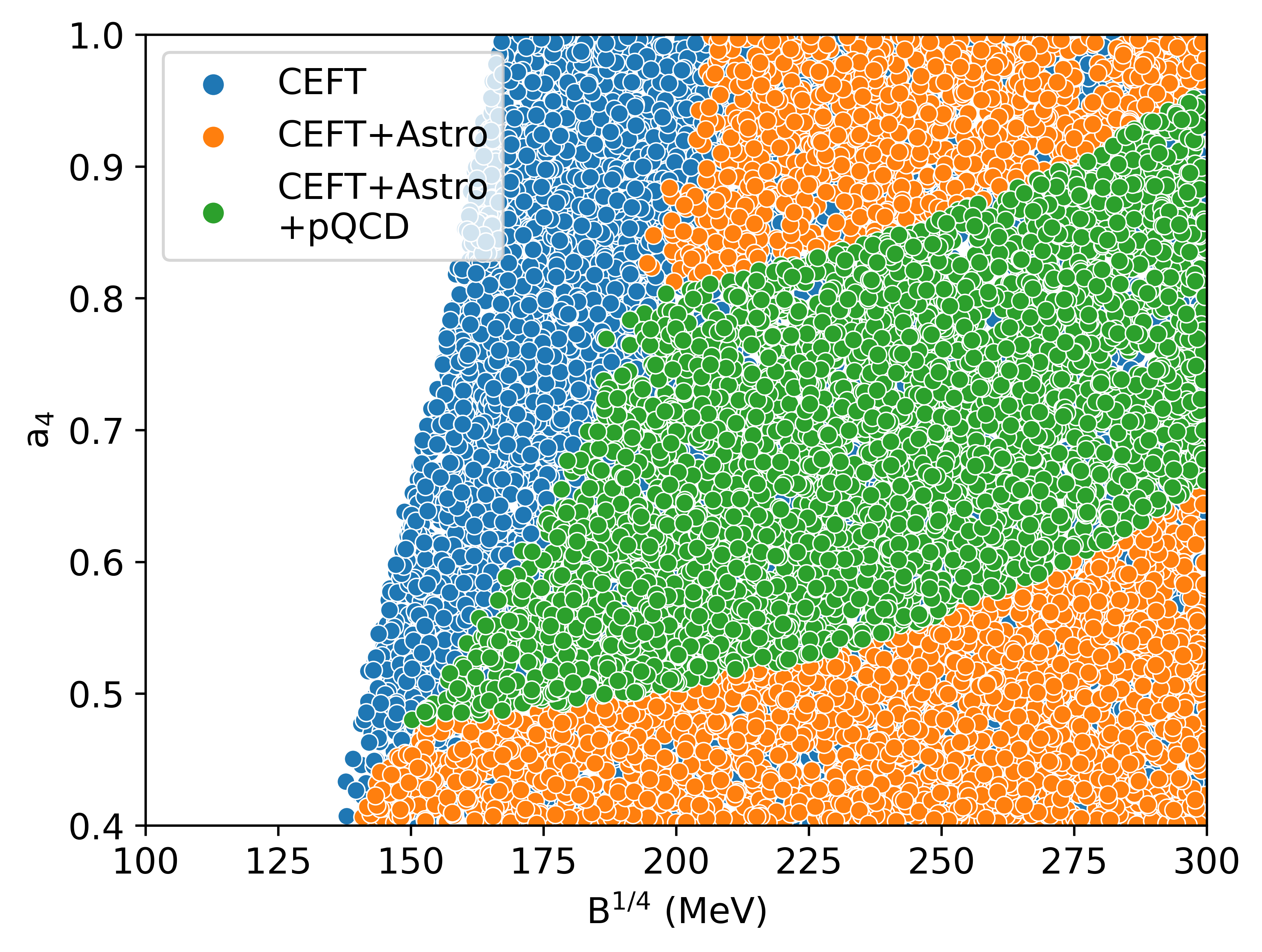}
    \caption{Scatter plot of $B$-$a_4$ parameters as a result of different filters. Same colour scheme as in Fig. \ref{fig:EoS} is used.}
    \label{fig:B-a_4scatter}
\end{figure}

We further investigate the constraints on the $B^{1/4}-a_4$ plane (Fig.~\ref{fig:B-a_4scatter}) upon imposition of pQCD constraints. No physical solutions are obtained for low values of $B$. The blue points represent the EoSs consistent with CEFT. The orange dots are obtained on the inclusion of astrophysical constraints. Imposing the pQCD constraint results in the green dots, further constraining the $B$-$a_4$ parameter space. We also note that this analysis puts a lower limit on the value of $a_4\approx0.48$. There is no strict upper limit on $a_4$ as it depends on the range of $B^{1/4}$ chosen. Similarly, we have a lower limit on the value of $B^{1/4}\approx135$ MeV purely from physical considerations. The inclusion of astrophysical and pQCD constraints raises this limit to around 150 MeV. 

\begin{figure}
    \epsscale{1.15}
    \plotone{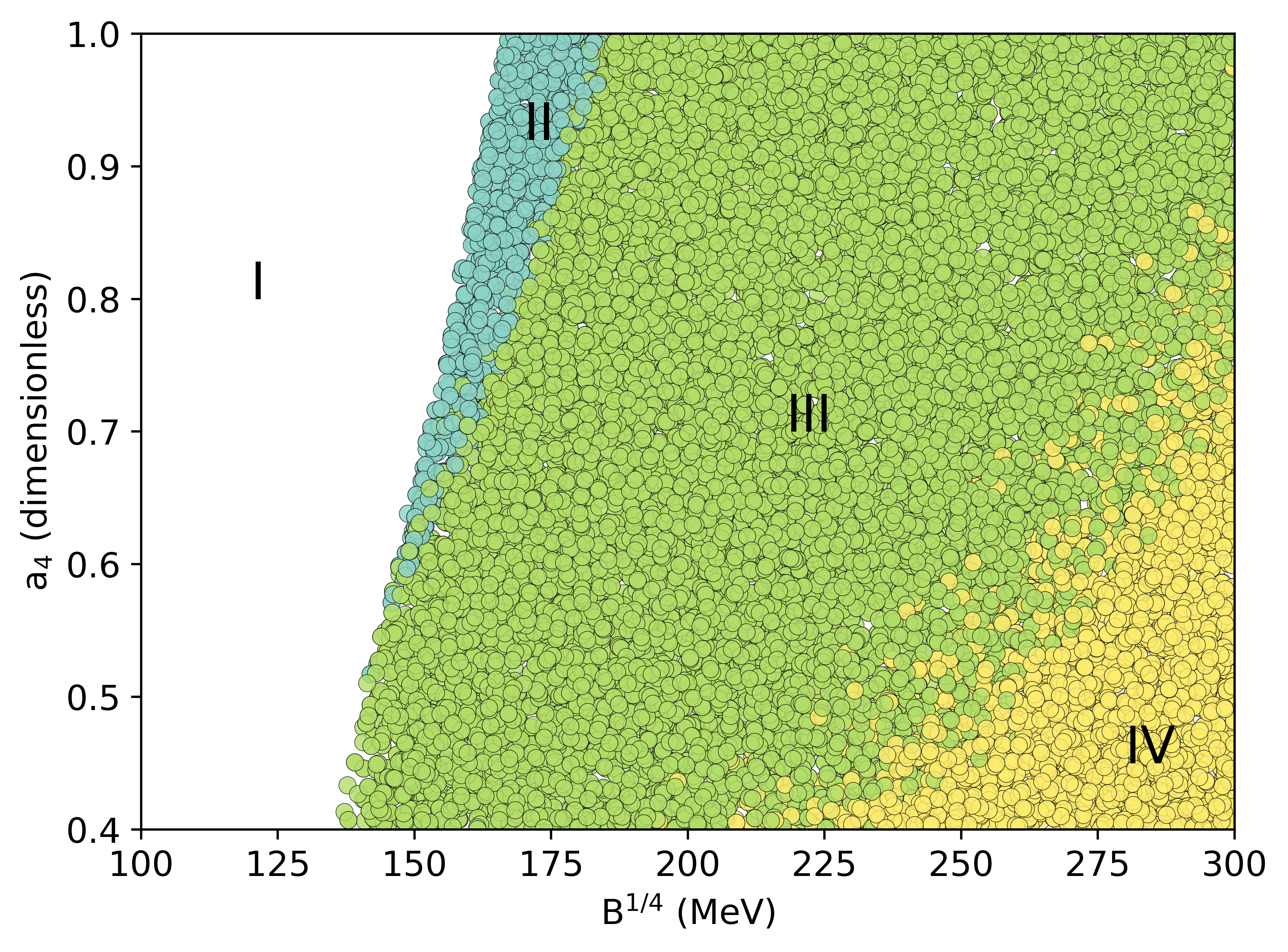}
    \caption{Different phases of matter realised in the maximum mass NS core across the $B$-$a_4$ plane (see text for details).}
    \label{fig:core_nature}
\end{figure}

In the scatter plot in Fig. \ref{fig:core_nature}, we analyse the phases of matter realised for maximum mass NSs, for the values of $B$-$a_4$ considered in Fig.~\ref{fig:B-a_4scatter}. For each EoS, we compare the critical densities for the mixed and pure quark phase with the central densities of maximum mass NS configuration for that EoS. For region IV, marked in yellow, the central densities are lower than the critical densities; hence, hybrid stars are not realized for such parameter values. For region III in green, central densities are higher than only the critical densities for the mixed phase, resulting in hybrid stars with only the mixed phase in the core. Region II in blue corresponds to hybrid stars whose cores have a pure quark phase. Only values of $a_4\gtrsim 0.5$ and $B^{1/4}\lesssim 185$ MeV result in a NS with pure quark core Region I corresponds to low values of $B$-$a_4$ for which no stable solutions for hybrid stars are obtained. A comparison of Fig. \ref{fig:core_nature} with Fig.~\ref{fig:B-a_4scatter} reveals that region III (hybrid stars with mixed phase cores) are allowed by combined filters, whereas region II (hybrid stars with pure quark phase in the core) are disfavoured by astrophysical constraints.

\section{Discussions}\label{sec:discussions}
In this work, we perform a systematic investigation of the role of EoS model parameters, both in the pure nucleonic and quark matter phases, in governing NS global observable properties. For this, we vary the model parameters (RMF model for the hadronic sector and MIT Bag model for the quark sector) within their present uncertainties to generate a flat prior within a Bayesian framework with a hard cut-off. We then impose constraints from CEFT (chiral effective field theory) at low densities, state-of-the-art astrophysical data from electromagnetic and gravitational wave observations at high densities, and pQCD (perturbative QCD) at very high densities to significantly restrict the allowed parameter space. We then use the posteriors to investigate possible correlations among the model parameters and with NS observables and extend the previous work \citep{ghosh2022multi} to include hadron-quark phase transition.

While the CEFT calculations constrain the EoS at low densities, NS astrophysical filters imposed constraints on the allowed parameter space for the nucleonic sector at higher densities. Although pQCD calculations are effective only at very high densities, which are not attained in NS interior, they significantly constrain the $B$-$a_4$ parameter space of the quark model and the high-density EoS. After applying the astrophysical and pQCD filters in addition to CEFT, our analysis showed that the maximum density reached inside NSs is 1.23 fm$^{-3}$ and the maximum baryon chemical potential reached is 1.97 GeV. We do not encounter any configuration with mass greater than 2.67 $M_{\odot}$. We also obtain a bound on tidal deformability $\Lambda_{1.4M_{\odot}} >$ 247. We get a fairly narrow range for the radius of 1.4 $M_{\odot}$ NS as $R_{1.4M_{\odot}} \in [11.4, 13.5]$ km. Among the bag model parameters we obtain a lower limit on $a_4 > 0.48$ and $B^{1/4} > 150$ MeV.

Next, we looked for physical correlations among model parameters and NS observables after applying different filters. First, after applying the CEFT constraints, we found a moderate correlation between $E_{sym}$ and $L_{sym}$. $m^*/m$ was strongly correlated with all the NS observables except $M_{max}$. $M_{max}$ also showed moderate correlation with $B^{1/4}$. Addition of astrophysical filters strengthened the correlation between $E_{sym}$ and $L_{sym}$ and between $m^*/m$ and $M_{max}$ making $m^*/m$ the most important parameter determining NS observables. The bag model parameters ($B^{1/4}$, $a_4$) did not show significant correlation with any other parameter or NS observable. However, note that there are also other observables dependant on quark content that should be studied \citep{alford2019signaturesads}. A moderate correlation between $n_{sat}$ and $R_{1.4M_{\odot}}$ and a weak correlation between $L_{sym}$ and $R_{1.4M_{\odot}}$ was seen in both cases. On applying all the filters, CEFT, astrophysical, and pQCD, we found a moderate correlation of $B^{1/4}$ with $M_{max}$ and $R$. The most important result of the study is that we find the emergence of correlation between the bag parameters $B^{1/4}$ and $a_4$ due to pQCD constraints. Other correlations remain unaffected.

We performed a detailed study of the bag model parameter space upon application of the different filters. We found that astrophysical observations favor higher values of bag constant $B$. Imposing the pQCD constraint along with astrophysical filters significantly restricts the $B$-$a_4$ parameter space. We also studied the phases of matter realized in the maximum NS configuration across the $B$-$a_4$ plane and found that although hybrid stars with mixed phase in the interior were allowed by some configurations, astrophysical observations disfavoured the existence of pure quark matter phase in the core. 

\subsection{Comparison with other works} \label{sec:comparison}
As outlined in the Introduction~\ref{sec:intro}, information from CEFT, multi-messenger astrophysical data, or pQCD has been used previously within interpolation or Bayesian schemes to constrain the EoS in hybrid stars. However, such studies mostly employed polytropic models or constant speed-of-sound parametrizations, which do not convey information about the underlying dense hadronic or quark matter. The advantage of our scheme is that one can directly probe the effect of each of these constraints on the EoS as well as on the physical correlations among the model parameters and NS observables.

In our investigation, we employed the RMF framework for the hadronic and MIT Bag model for the quark matter sector. Although few studies have been performed for hybrid stars using such models, selected RMF parametrizations were used to conclude which of them are compatible with current observations \citep{parisi2021hybrid, nandi2018, nandi2021}. In contrast, we perform a systematic investigation spanning the entire allowed parameter space (which includes such parametrizations), which allows us to investigate the role of each nuclear or quark matter parameter in governing the global NS properties through correlation studies.

The \textbf{range of posteriors in $M_{max}$ and $R_{1.4M_{\odot}}$} resulting from the different filter functions (see Table~\ref{table:densityminmax}) can be compared to those in Table. 1 of \citet{annala2018gwconstraints} where only polytropes were used. \citet{kurkela2014constraining} did not realise any configuration above 2.5 $M_{\odot}$. Our results are consistent with those obtained by~\citet{annala2018gwconstraints} (2.0 - 2.7 $M_{\odot}$). The \textbf{range for $R_{1.4M_{\odot}}$} we obtain (11.4-13.5 km) is narrower compared to previous works that use only polytropic EoSs. \citet{hebeler2013} obtained a range for $R_{1.4M_{\odot}}$ as 10-13.7 km applying CEFT and maximum mass constraint of 1.97 $M_{\odot}$. \citet{kurkela2014constraining} imposed pQCD constraint with a mass threshold of 2 $M_{\odot}$ to obtain a slightly shifted range of 11-14.5 km. \citet{annala2018gwconstraints} further imposed the constraint of tidal deformability of GW170817 and obtained a range of 10.7-13.6 km (one should note that the upper limit on $\Lambda_{1.4M_{\odot}}$ used there is 800, instead of 720 used here). In our work, we find the highest lower bound of $R_{1.4M_{\odot}} > 11.4$ km. Employing the Bag model and 3 RMF EoSs, \citet{nandi2018} obtained an upper limit of $R_{1.4M_{\odot}}<13.5$ km (for  $\Lambda_{1.4M_{\odot}} < 800$), which was later updated to include 28 RMF EoSs \citep{nandi2021} to obtain an improved limit of $R_{1.4M_{\odot}}\lesssim13.3$ km (for $\Lambda_{1.4M_{\odot}} < 720$), consistent with the range we obtain in this work. \citet{traversi2020} performed a similar study using the RMF model in the Bayesian framework, varying five of the nuclear empirical parameters for different types of priors and found hints of a phase transition to a chiral symmetry restored phase predicting $R_{1.4M_{\odot}} \sim 12$ km. 
Another recent study \citep{huth2022}, similar to \citet{ghosh2022multi}, performed Bayesian analysis combining information from nuclear theory, experiments and astrophysical observations, obtaining a radius estimate of $R_{1.4M_{\odot}}=11.93^{+0.39}_{-0.41}$ km. 
Although these studies do not consider an explicit phase transition in their model, their results are in agreement with the range of radius we obtain.
\citet{Most2018radiuslamfromgw} imposed multi-messenger constraints on parametrized EoSs, by modeling the phase transition with a jump in energy density leading to a separate twin branch, a new family of stars. 
NICER observations provide simultaneous measurement of the mass and radius of NSs. Analysis of NICER data for PSR J0030+0451 yields equatorial radius of $R_{e}=13.02^{+1.24}_{-1.06}$ km \citep{miller2019nicermrj0030ads} for a mass of $\sim 1.44 M_{\odot}$ consistent with an independent study \citep{riley2019nicermrj0030ads} which reported $R_{e}=12.71^{+1.14}_{-1.19}$ km. Analysis of GW170817 binary neutron star merger data also generates a mass-radius posterior obtained for a low-spin prior \citep{AbbottPRL2018gw170817massradius}. The final posterior set obtained passes through $1\sigma$ contours of mass-radius distributions of both of these studies. Study of another pulsar PSR J0740+6620 (the one with maximum mass) done by \citet{miller2021nicermrj0740ads} and \citet{riley2021nicermrj0740ads} obtain radius measurements as $R_{e}=13.7^{+2.6}_{-1.5}$ km and $R_{e}=12.39^{+1.30}_{-0.98}$ km respectively for $\sim2M_{\odot}$ NS. The posteriors satisfy $2\sigma$ contours of the joint mass-radius distribution of this pulsar. Results of our investigation also rule out \textbf{tidal deformability values} below 247 ($\Lambda_{1.4M_{\odot}} > 247$). In comparison, \citet{annala2018gwconstraints} had obtained a much lower limit ($\Lambda_{1.4M_{\odot}} > 120$).

\begin{figure}
    \epsscale{1.15}
    \plotone{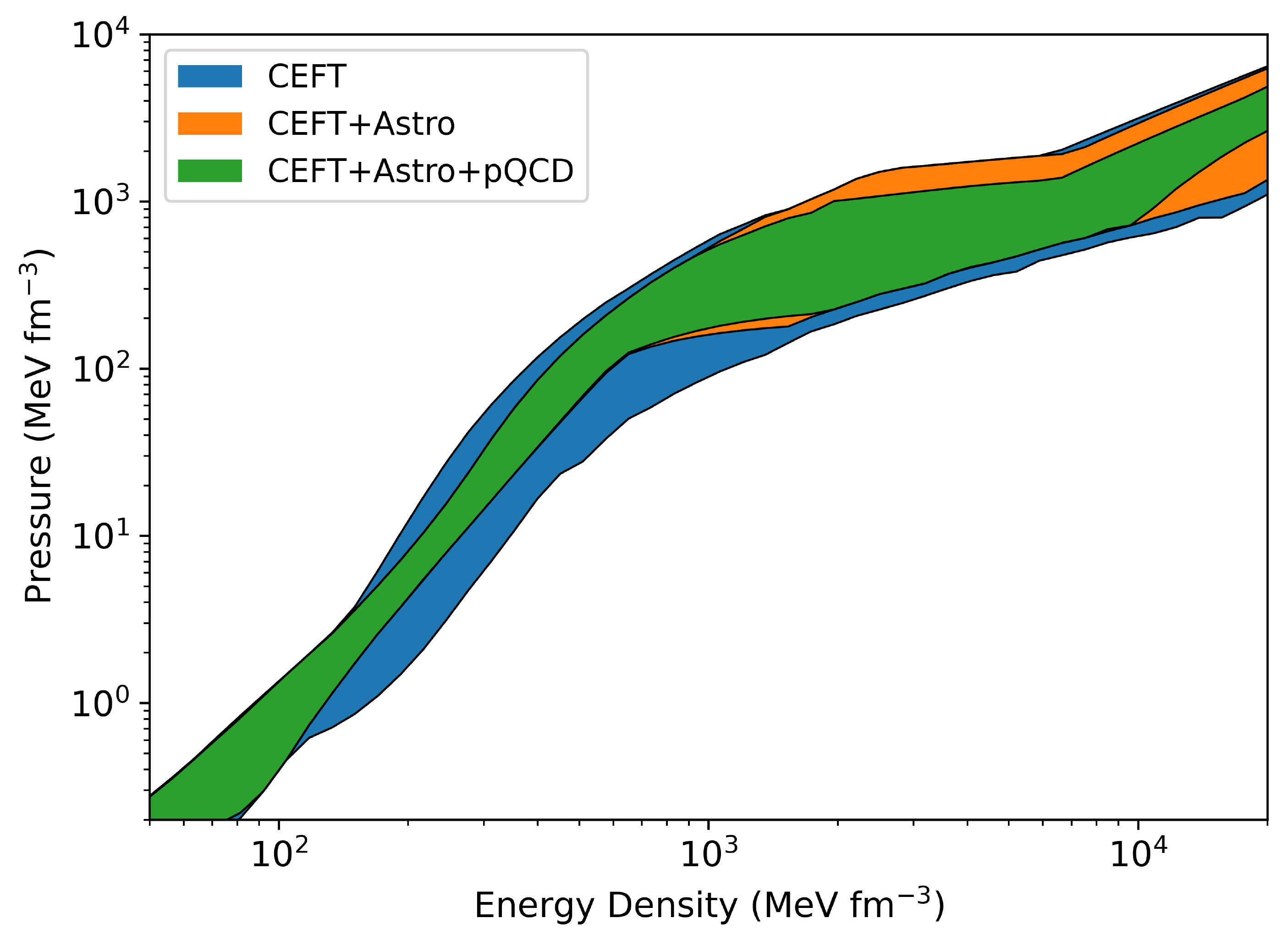}
    \caption{EoS (pressure vs energy density) contours corresponding
    to the EoS contours shown in Fig.~\ref{fig:EoS}}
    \label{fig:eos2}
\end{figure}

Fig~\ref{fig:eos2} shows the posterior EoS bands as shown in Fig.~\ref{fig:EoS} with pressure as a function of energy density. We confirm that the CEFT constraints are most important for low energy densities. In the energy density range 100-1000 MeV fm$^{-3}$, the astrophysical constraints are most constraining. pQCD effects are effective at very high densities corresponding to energy density $> 10000$ MeV fm$^{-3}$. We find a kink in the EoS at around 600-700 MeV fm$^{-3}$, indicating the onset of the hadron-quark phase transition. This was recently proposed as strong evidence for the existence of quark matter (see Fig. 10 of \citet{kurkela2014constraining}, Fig. 3 of \citet{annala2018gwconstraints}, Fig. 1 of \citet{annala2020evidence}).

For a fixed RMF EoS (NL3$\omega\rho$), ~\citet{nandi2018} reported upper limits of 163 MeV for $B^{1/4}$ and 0.65 for $a_4$. Varying all the model parameters within their present uncertainties and imposing multi-physics and multi-messenger constraints, we obtain a lower limit of $a_4>0.48$. From our analysis of Fig.~\ref{fig:core_nature}, 
for most of the bag parameter values, NSs with a mixed core are realized (region III), while a pure quark core is formed for a narrow band represented by region II. This is in contrast with the findings of \citet{nandi2021}, which reported a mixed phase core for most of the hybrid stars but also several EoSs with pure quark cores. Our results are however consistent with the findings of ~\citet{weissenborn2011quark}, which studied hybrid stars considering two RMF EoSs (TM1 and NL3) and found pure quark cores for a very small parameter range. They concluded that a pure quark core appears if the phase transition begins around nuclear saturation density which happens in the case of low $B^{1/4}$ and high $a_{4}$.

\subsection{Implications and future directions}
This work is the first systematic study of the application of combined filters from nuclear physics calculations, astrophysical data, and perturbative QCD to restrict the range of model parameters in hybrid stars for realistic EoSs. The results of this study clearly identify the effect of each of the filters on the underlying empirical parameters and correlations, as well as the role of the parameters in governing NS observables. This could help significantly reduce the parameter space and, therefore, the computational time for parameter estimation for future gravitational wave searches. The analysis also imposes strong restrictions on the allowed parameter space of the widely used bag model using the combined filters and also draws important conclusions about the possibility of finding quark matter in the interior of hybrid stars.

From this work on NS astrophysical observables, it would be interesting to investigate the consequences of the restricted parameter space (particularly for the bag model). Our allowed parameter space disfavors hybrid stars with pure quark phase but allows mixed phase in the interior. We leave this for a possible future investigation.

\section*{Acknowledgements}
The authors acknowledge usage of the IUCAA HPC computing facility for the numerical calculations. S.S. would like to thank Bikram Keshari Pradhan and Dhruv Pathak for useful discussions during this work.

\bibliography{main}{}
\bibliographystyle{aasjournal}

\end{document}